\documentclass[11pt,twoside]{article}
\usepackage{asp2010}
\resetcounters

\begin{document}

\title{Magnetic Cataclysmic Variables Discovered in the Palomar Transient Factory}
\author{Bruce Margon,$^1$ David Levitan,$^2$ Thomas A. Prince,$^2$ Gregg Hallinan,$^2$ and the PTF Collaboration
\affil{$^1$ Department of Astronomy and Astrophysics, University of California, 1156 High Street, Santa Cruz, CA 95064, USA.}
\affil{$^2$ Division of Physics, Mathematics, and Astronomy, California Institute of Technology, Pasadena, CA 91125, USA.}}

\begin{abstract}
The Palomar Transient Factory proves to be a prolific source of Magnetic Cataclysmic Variables, selected by their distinctive photometric variability, and followed up spectroscopically.
Here, we present six new candidate systems, together with preliminary photometric periods and spectra.
\end{abstract}

\section{Introduction}
Using the Palomar Transient Factory (PTF) synoptic survey, we have discovered via large amplitude optical variability, followed by spectroscopy, six previously unreported magnetic cataclysmic variables (CVs). As most such objects have previously been X-ray selected, the availability of wide-area synoptic surveys such as the PTF marks the start of a systematic technique to select magnetic CVs optically.  Despite extremely strong He$\;${\small \uppercase\expandafter{\romannumeral 2}} optical emission in all the spectra, indicating the presence of very high excitation gas, only two of the six are reported X-ray sources.  Thus the possibility of significant (X-ray) selection bias in previous samples is clear. We anticipate that the PTF and related future surveys will invigorate the study of magnetic CVs by providing a large, homogeneously-selected sample for further study.

\section{Magnetic Cataclysmic Variables}

Magnetic cataclysmic variables are close red dwarf/white dwarf binary systems that can provide unique insights on the late stage evolution of mass exchange binaries with compact companions.  These systems typically have orbital periods of 1.5--6\,h, and show sporadic optical outbursts typical of other subclasses of cataclysmic variables (CVs). As the name implies, the white dwarf is strongly magnetic, with field strengths of a few through tens of MG for the so-called Intermediate Polars, and yet more intense fields for polars (``AM Her stars''). As in the non-magnetic CVs, there is strong accretion flow and large amplitude photometric outbursts, but in addition, the magnetic field dramatically disrupts and controls the accretion flow.  Magnetic CVs therefore can be laboratories for the detailed study of accretion into deep potential wells in the presence of strong fields, a scenario which comes up again and again in a variety of astrophysical objects.  Further, these objects typically exhibit a range of unique observational signatures which allow the systems to be characterized in great detail.  Often both the visible and X-ray light curves will show the binary orbital period, the white dwarf spin period, eclipses, as well as moderate and extreme outbursts, both roughly periodic and aperiodic, variable optical polarization, and cyclotron harmonic features in the visible spectrum. There are currently dozens of such systems, although not all simultaneously exhibit all of the observational diagnostics noted above.

\section{The PTF and Magnetic CVs}

The majority of magnetic CVs have thus far been discovered via their strong X-ray emission, although there are certainly a few prominent optically-discovered exceptions, for example the old novae DQ Her and GK Per, as well as some previously-cataloged variable stars.  However, recently the Palomar Transient Factory\footnote{\url{http://www.astro.caltech.edu/ptf}} \citep[PTF; ][]{2009PASP..121.1395L,2009PASP..121.1334R} has provided the capability to optically discover and characterize a very large number of highly variable objects.  The PTF uses the Oschin 48-inch Schmidt telescope at the Palomar Observatory to image $7.2\,\mathrm{deg}^2$ in each exposure to a depth of $R\sim20.6$, with a typical cadence of 1--5\,d.  The PTF is particularly suited for the discovery of CVs due to its wide sky coverage ($\mathord{\sim}16,000\,\mathrm{deg}^2$), excellent photometric calibration, and broad magnitude range, which covers both the brightest and faintest currently known magnetic CVs. Subsets of variable objects are then followed up spectroscopically to determine their nature. 

\begin{figure}
{\centering
\plotone{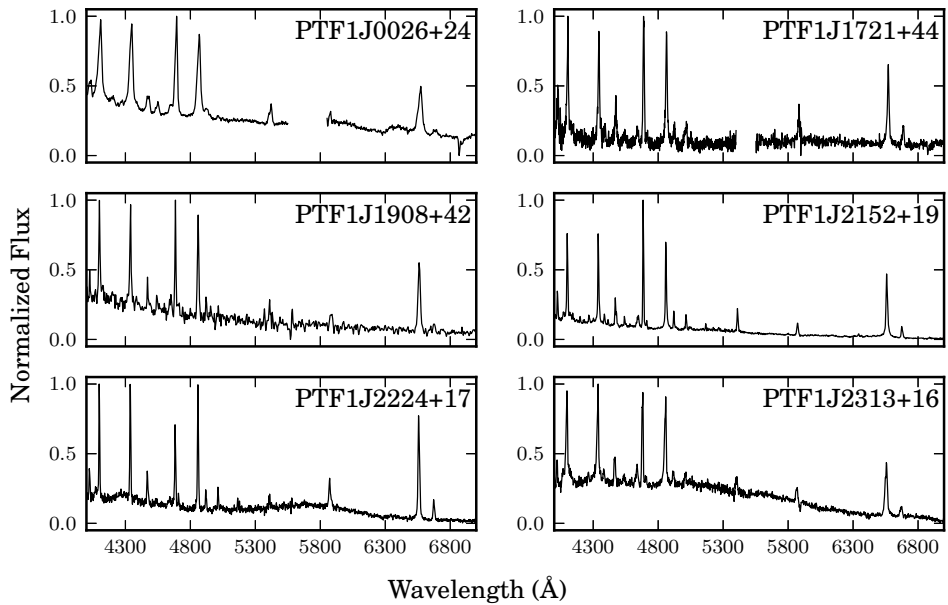} }
\caption{Spectra obtained at the Keck, Palomar 5-m and Lick 3-m telescopes of six newly discovered magnetic CVs, selected via large amplitude photometric variability in the PTF.  All exhibit intense Balmer and He$\;${\small \uppercase\expandafter{\romannumeral 2}} emission.  Note also the extreme strength of He$\;${\small \uppercase\expandafter{\romannumeral 2}} $\lambda4686$ relative to H$\beta$ and H$\alpha$, a signature of Magnetic CVs.}
\label{fig:spectroscopy}
\end{figure}

In this followup process we have recently discovered a group of six highly variable PTF objects which typically vary in the $R\sim17\mbox{--}21$ range and prove to have the distinctive spectra of magnetic CVs. In Figure \ref{fig:spectroscopy} we show the spectra of these objects, which are dominated by intense Balmer and He$\;${\small \uppercase\expandafter{\romannumeral 2}} emission lines. Most remarkably and unusually, in all the spectra He$\;${\small \uppercase\expandafter{\romannumeral 2}}  $\lambda4686$ equals or exceeds the strength of H$\beta$ and H$\alpha$.  This extraordinary condition is one rather unambiguous spectral diagnostic of magnetic CVs.  

\begin{figure}
{\centering \plotone{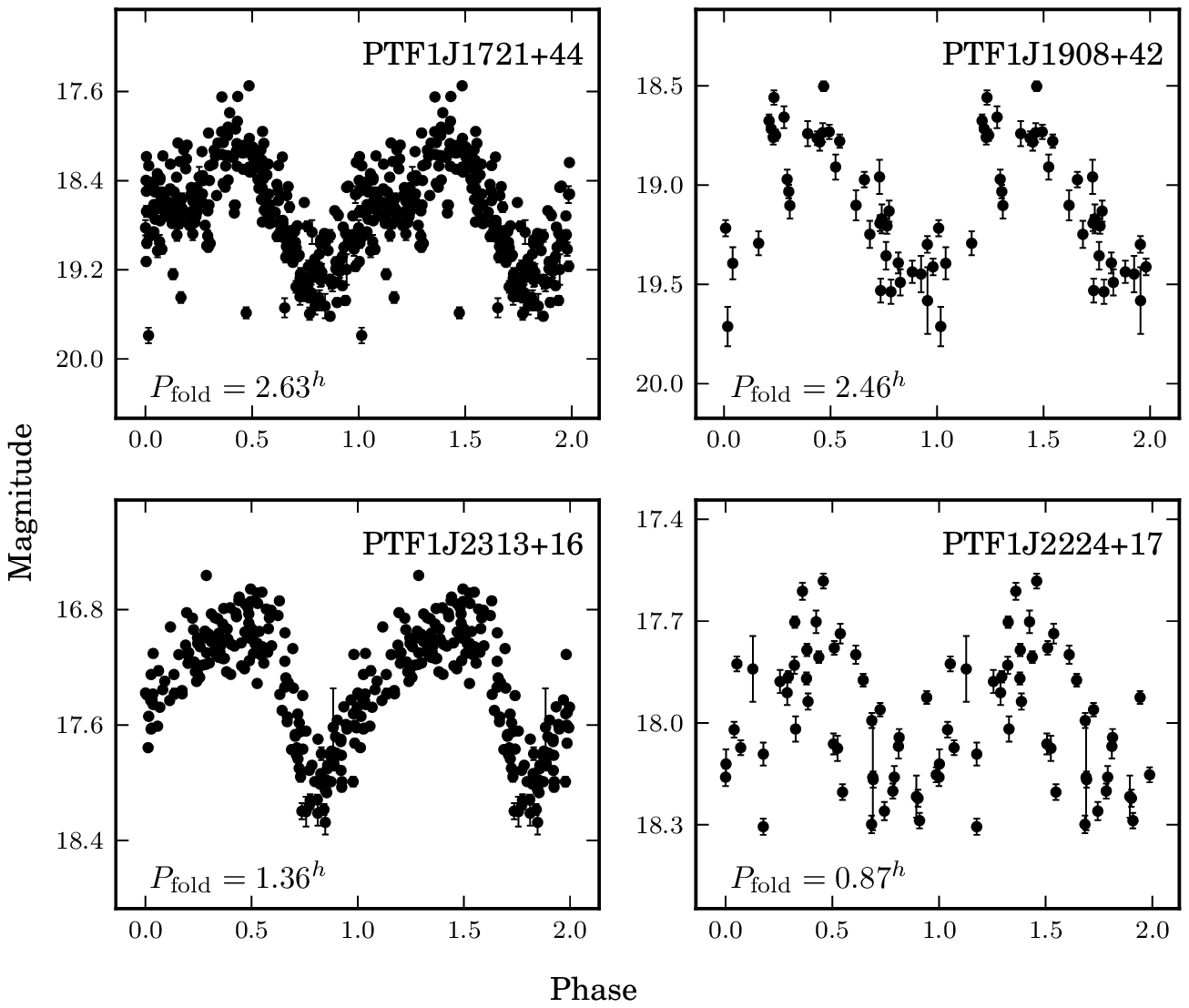}}
\caption{Folded PTF data for those systems that exhibit high confidence periodic modulation. The best-fit periods are indicated for each system.}
\label{fig:foldlc}
\end{figure}

\begin{figure}
{ \centering
\plotone{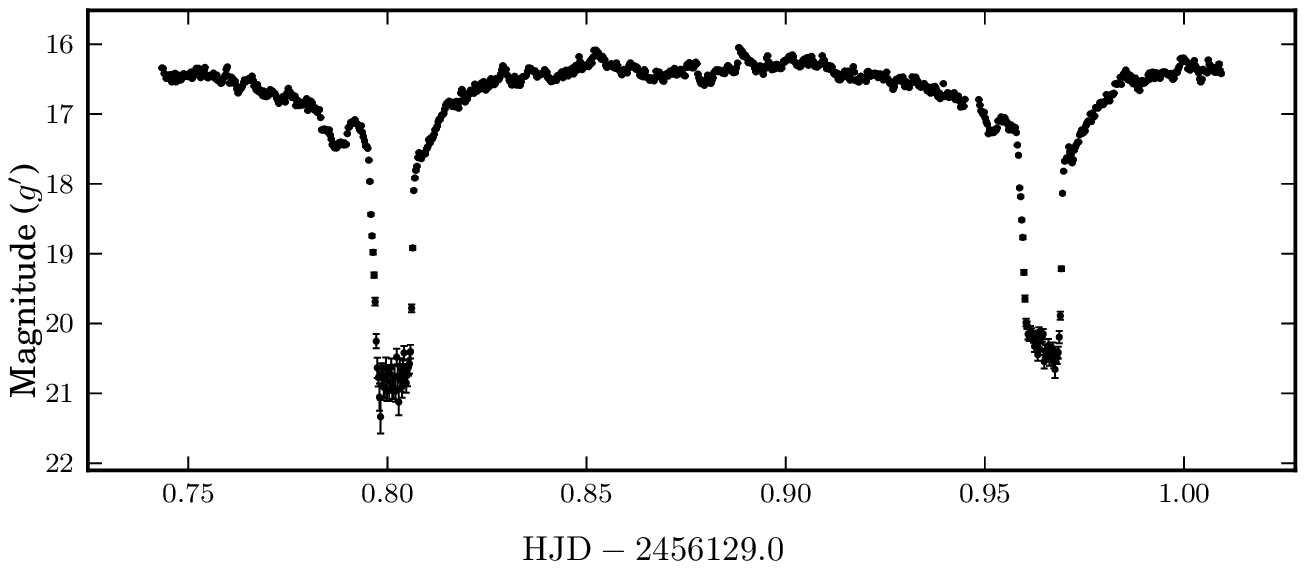}
}
\caption{Photometry of PTF1J2152+19, one of the newly discovered magnetic CVs, obtained with the Lick 3-m Shane telescope in July 2012.  Note the very deep ($\mathord{\sim}5\,$mag) eclipse; there are only a small number of fully eclipsing magnetic CVs known. From many observed orbital cycles, we derive a period of 3.91682\,h.}
\label{fig:1221e}
\end{figure}

Further, for 5 of the 6 objects we have also discovered photometric periods, using data from the PTF and other synoptic surveys (Figure \ref{fig:foldlc}), as well as newly obtained observations from the Lick Shane telescope.  In 4 of these 5 cases, our measured periods are in the 1.4--3.9\,h range and are undoubtedly orbital; the 5th case shows a 52\,m period, which is presumptively the white dwarf spin period.
In Figure \ref{fig:1221e}, we show one particularly well-studied system, a newly discovered magnetic CV with a 3.92\,h orbital period and a deep ($\mathord{\sim}5\,$mag) eclipses; this object joins the rare subclass of deeply eclipsing magnetic CVs, and is thus especially amenable to a total system solution.

Without the detection of polarization, and/or observation of both orbital and white dwarf spin periods simultaneously in the same object, the classification as magnetic CVs and further
sub-classification as polars or intermediate polars must remain tentative; one or more of these objects could also prove to be SW Sex stars, for example. We do have very recent polarimetric observations on hand, which when reduced may address this issue.

\section{X-ray Emission, or Lack Thereof}

The lack of appearance of most of these PTF optically-selected magnetic CVs in previous X-ray surveys is interesting.  We believe that two of the six stars, PTF1J1721+44 and PTF1J2313+16, are the previously unrecognized optical counterparts of otherwise unremarkable X-ray sources in the ROSAT All Sky Survey and the XMM-Newton Slew Survey, respectively, as the positional coincidences are excellent, but the remaining 4 objects are to our knowledge undetected by any X-ray mission. This cannot simply be the result of distance, as all six PTF objects are of comparable median optical magnitudes, nor interstellar photoelectric absorption, as all the objects are at high galactic latitude.  If only one or two objects were undetected, we could attribute this to chance, or the source temporarily lying in an extreme low state.  However, with 4 out of the 6 systems currently not known to be X-ray emitters, and the sample gathered from the first homogenous set of optically rather than X-ray selected sources, there is an obvious suspicion that the properties of optically and X-ray selected magnetic CVs might be intrinsically different.   

\bibliography{aspauthor}

\end{document}